\documentclass{emulateapj}
\usepackage{natbib, aas_macros}
\usepackage{color}
\usepackage{lscape}

\begin{document}

\title{The Quasar-LBG two-point angular cross-correlation function at $z \sim$ 4 \\ in the COSMOS field}
\author{H. Ikeda,\altaffilmark{1,2,3,4} T. Nagao,\altaffilmark{5} Y. Taniguchi,\altaffilmark{5} K. Matsuoka,\altaffilmark{2,6,7} M. Kajisawa,\altaffilmark{1,5} M. Akiyama,\altaffilmark{8} T. Miyaji,\altaffilmark{9,10} N. Kashikawa,\altaffilmark{11,12} T. Morokuma,\altaffilmark{13} Y. Shioya,\altaffilmark{5}  M. Enoki,\altaffilmark{14} P. Capak,\altaffilmark{15} A. M. Koekemoer,\altaffilmark{16} D. Masters,\altaffilmark{15,17} M. Salvato,\altaffilmark{18} D. B. Sanders,\altaffilmark{19} E. Schinnerer,\altaffilmark{20} and N. Z. Scoville\altaffilmark{15} }
\altaffiltext{1}{Graduate School of Science and Engineering, Ehime University, Bunkyo-cho, Matsuyama 790-8577, Japan; email: ikeda@cosmos.phys.sci.ehime-u.ac.jp}
\altaffiltext{2}{Research Fellow of the Japan Society for the Promotion of Science}
\altaffiltext{3}{Institute of Astronomy and Astrophysics, Academia Sinica, P.O. Box 23-141, Taipei 10617, Taiwan}
\altaffiltext{4}{Astronomy Data Center, National Astronomical Observatory of Japan, 2-21-1 Osawa, Mitaka, Tokyo 181-8588, Japan}
\altaffiltext{5}{Research Center for Space and Cosmic Evolution, Ehime University, Bunkyo-cho, Matsuyama 790-8577, Japan}
\altaffiltext{6}{Department of Physics and Astronomy, Seoul National University, 599 Gwanak-ro, Gwanak-gu, Seoul 151-742, Korea}
\altaffiltext{7}{Department of Astronomy, Graduate School of Science, Kyoto University, Kitashirakawa-Oiwake-cho, Sakyo-ku, Kyoto 606-8502, Japan}
\altaffiltext{8}{Astronomical Institute, Tohoku University, 6-3 Aramaki, Aoba-ku, Sendai 980-8578, Japan}
\altaffiltext{9}{Instituto de Astronom\'ia, Universidad Nacional Aut\'onoma de M\'exico, Ensenada, Baja California, Mexico}
\altaffiltext{10}{University of California, San Diego, Center for Astrophysics and
Space Sciences, 9500 Gilman Drive, La Jolla, CA 92093-0424, USA}
\altaffiltext{11}{Optical and Infrared Astronomy Division, National Astronomical Observatory, Mitaka, Tokyo 181-8588, Japan}
\altaffiltext{12}{Department of Astronomy, School of Science, Graduate University for Advanced Studies, Mitaka, Tokyo 181-8588, Japan}
 \altaffiltext{13}{Institute of Astronomy, Graduate School of Science, University of Tokyo, 2-21-1 Osawa, Mitaka 181-0015, Japan}
\altaffiltext{14}{Faculty of Bussiness Administration, Tokyo Keizai University, 1-7-34 Minami-cho, Kokubunji, Tokyo 185-8502, Japan}
\altaffiltext{15}{California Institute of Technology, MC 105-24, 1200 East California Boulevard, Pasadena, CA 91125, USA}
 \altaffiltext{16}{Space Telescope Science Institute, 3700 San Martin Drive, Baltimore, MD 21218, USA}
\altaffiltext{17}{Department of Physics and Astronomy, University of California, 900 University Ave, Riverside, CA 92521, USA}
 \altaffiltext{18}{Max-Planck-Institut f\"ur Plasmaphysik, Boltzmanstrasse 2, D-85741 Garching, Germany}
 \altaffiltext{19}{Institute for Astronomy, 2680 Woodlawn Drive, University of Hawaii,
Honolulu, HI 96822, USA}
\altaffiltext{20}{Max-Planck-Institut f\"ur Astronomie, K\"onigstuhl 17, D-69117 Heidelberg, Germany}
\shortauthors{Ikeda et al.}    

\begin{abstract}
 In order to investigate the origin of quasars, we estimate the bias factor for low-luminosity quasars at high redshift for the first time.
 In this study, we use the two-point angular cross-correlation function (CCF) for both low-luminosity quasars at $-24<M_{\rm 1450}<-22$ and Lyman-break galaxies (LBGs).
Our sample consists of  both 25 low-luminosity quasars (16 objects are spectroscopically confirmed low-luminosity quasars)
in the redshift range $3.1<z<4.5$ and 835 color-selected LBGs with $z^{\prime}_{\rm LBG}<25.0$ 
at $z\sim4$ in the COSMOS field.
We have made our analysis for the following two quasar samples;
(1) the spectroscopic sample (the 16 quasars confirmed by spectroscopy), and
(2) the total sample (the 25 quasars including 9 quasars with photometric redshifts).
The bias factor for low-luminosity quasars at $z\sim4$ is derived by utilizing the
quasar-LBG CCF and the LBG auto-correlation function. 
We then obtain the $86\%$ upper limits of the bias factors for low-luminosity quasars, that are 5.63 and 
10.50 for the total and the spectroscopic samples, respectively.
These bias factors correspond to the typical dark matter halo masses, 
log $(M_{\rm DM}/(h^{-1}M_{\odot}))=$$12.7$ and $13.5$, respectively. 
This result is not inconsistent with the predicted bias for quasars which is estimated 
by the major merger models.
 \end{abstract}

\section{Introduction}
The observed close relationship between the mass of the spheroidal component of a galaxy and its central supermassive black hole (SMBH) suggests that the evolution of galaxies and SMBHs are closely related (e.g., \citealt{2003ApJ...589L..21M, 2004ApJ...604L..89H, 2013ApJ...764..184M}). Accordingly some important questions arise as when and how such a co-evolution has been established and what physical processes are essentially important. A straightforward approach to explore these issues is investigating the statistical properties of quasars at high redshifts, where the quasar activity (that corresponds to the growth of SMBHs) is much more active than in the local universe (e.g., \citealt{2006AJ....131.2766R,2009MNRAS.399.1755C}), because different evolutionary scenarios for SMBHs predict different statistical properties of quasars as function of redshift (e.g., \citealt{2007ApJ...662..110H}).

The quasar activity is thought to be powered by mass accretion onto a SMBH at the center of massive galaxies (\citealt{1984ARA&A..22..471R}). The most efficient gas fueling mechanism is thought to be major and minor mergers of galaxies (e.g., \citealt{1988ApJ...325...74S,1999ApJ...524...65T}; see also \citealt{2013ASPC..477..265T}). Therefore, in order to constrain the triggering mechanism for quasar activity, we need detailed studies of environmental properties of quasars. 

Motivated by these considerations, the two-point auto-correlation function (ACF) of quasars has been studied based on wide-field survey data, e.g., 2dF Quasar Redshift Survey and Sloan Digital Sky Survey \citep[e.g.,][]{2005MNRAS.356..415C,2006MNRAS.371.1824P,2006ApJ...638..622M,2007ApJ...658...85M,2007AJ....133.2222S,2008MNRAS.383..565D,2009ApJ...697.1656S,2009ApJ...697.1634R,2010MNRAS.409.1691I,2012MNRAS.424..933W}. The ACFs have also been investigated for active galactic nuclei (AGNs) including quasars selected through X-ray data \citep[e.g.,][]{2007ApJS..172..396M, 2008ApJS..179..124U,2009A&A...494...33G,2011ApJ...741...15S, 2012ApJ...746....1K,2013MNRAS.428.1382K,2014ApJ...796....4A}. Most of these studies have shown that luminous AGNs tend to live in massive dark matter halos ($\sim10^{12}-10^{13.5} h^{-1}M_{\odot}$), suggesting that luminous AGN activity is triggered by galaxy mergers because such massive dark matter halos could be assembled by successive mergers of small dark matter halos. It is also reported that galaxy mergers do not account for the majority of the moderate X-ray luminous AGNs at $z\lesssim2.2$ (\citealt{2011ApJ...736...99A}). However, it is difficult to investigate small-scale clustering properties of quasars because of their  low number density.

Here it should be noted that the mass of dark-matter halos hosting quasars can be estimated also through the two-point cross-correlation function (CCF) for quasars and galaxies, combined with the ACF for galaxies. The advantage of the CCF is that the required size of the quasar sample is relatively smaller than the ACF analysis, though we need enough number of galaxies around quasars for the CCF analysis. In addition, it is possible to study the small-scale clustering properties of quasars through the CCF, that is too challenging for the ACF study. Therefore, in order to understand the triggering mechanism of quasar activity, it is useful to investigate the CCF for quasars and galaxies around them. Several pioneering studies have been made to date  \citep[e.g.,][]{2005ApJ...627L...1A, 2009MNRAS.397.1862P,2009MNRAS.394.2050M,2009ApJ...701.1484C,2011ApJ...726...83M,2011ApJ...731..117H,2013ApJ...773..175Z, 2011PASJ...63S.469S, 2012MNRAS.420..514M, 2013ApJ...775...43K, 2013ApJ...778...98S, 2015arXiv150103898K}. For instance, Zhang et al. (2013) investigated the spatial clustering of galaxies around quasars at redshifts from 0.6 to 1.2. They found that the clustering amplitude is significantly larger for quasars with more massive black holes, or with bluer colors, while 
there is no dependence on quasar luminosity. This suggests that the mass of dark matter halos  in which quasars reside is not correlated with the quasar luminosity. In addition, it is possible that the triggering mechanism of high- and low-luminosity quasars may be the same in this luminosity range.

The CCF of quasars and galaxies has also been investigated at higher redshifts. Shirasaki et al. (2011) investigated the projected CCF of AGNs and galaxies at redshifts from 0.3 to 3.0 and found significant excess of galaxies around the AGNs.
They found that AGNs at higher redshifts reside in denser environments than those at lower
redshifts. This suggests that major mergers are the preferred mechanism to trigger AGN activity at high redshifts. They also reported that there is no luminosity dependence of AGN clustering. 
At $z\sim3$, \cite{2008ApJ...673L..13F} studied the two-point angular CCF of AGNs and Lyman break galaxies (LBGs)  \citep[see also][]{2011MNRAS.414....2B} and found that AGNs tend to be clustered  more strongly than LBGs.  They also found no luminosity dependence of AGN clustering. At $z>3$, clustering of LBGs around quasars has been recently studied \citep[e.g.,][]{2013MNRAS.432.2869H}. However the luminosity dependence of quasar clustering is not studied, due to the lack of adequate samples of low-luminosity quasars and galaxies around them. Therefore, the triggering mechanism of low-luminosity quasars at $z>3$ has not yet been studied so far.

 Motivated by the situation described above, we study the two-point angular CCF of the 25 low-luminosity quasars (16 objects are spectroscopically confirmed low-luminosity quasars) and 835 LBGs in the redshift range $3.1<z<4.5$ in the COSMOS field. Then we derive the bias factor for low-luminosity quasars and constrain the dark matter halo mass in which low-luminosity quasars at $z\sim4$ exist. The outline of this paper is as follows.
In Section 2, we describe the data of low-luminosity quasars focused in this study and the method used for the photometric selection of LBG candidates.
In Section 3, we report the results of the clustering analysis of quasars and LBGs. 
In Section 4 and 5, we give our discussion and summary. Throughout this paper we adopt a $\Lambda$CDM cosmology with $\Omega_m$ = 0.3,\  $\Omega_{\Lambda}$ = 0.7, $\sigma_{\rm 8}$ = 0.9, and a Hubble constant of $H_0$ = 70 km s$^{-1}$ Mpc$^{-1}$. 
We use the AB magnitude system. All of the errors reported in this paper are 1 sigma.

\section{The Sample}

\subsection{The Cosmic Evolution Survey}
Wide and deep multi-wavelength data are publicly available in the Cosmic Evolution Survey (COSMOS) field
(Scoville et al. 2007). Therefore we can select large numbers of high-$z$ galaxies and quasars in the same field to study their statistical properties such as the CCF. Another advantage of the COSMOS dataset is the dense sampling in the optical wavelength with intermediate-band filters, that makes the estimates of the photometric redshifts of galaxies far more accurate than in other deep-survey fields (\citealt{2009ApJ...690.1236I}; see also \citealt{2010ApJS..189..270C}). For the above reasons, we decided to focus on the COSMOS field for the CCF analysis at $z\sim4$. 

COSMOS is a treasury program of the Hubble Space Telescope (HST). It comprises 270 and 320 orbits allocated in the HST Cycles 12 and 13, respectively (\citealt{2007ApJS..172....1S}; Koekemoer et al. 2007).
The COSMOS field covers an area of $\sim\rm 1.4^{\circ} \times 1.4^{\circ}$ square which corresponds to $\sim2\deg^2$, centered at R.A. (J2000) = 10:00:28.6 and Dec. (J2000) =+02:12:21.0. We use an upgraded version of the photometric redshift catalogue from \cite{2009ApJ...690.1236I}  (see also \citealt{2007ApJS..172...99C}) including the new UltraVISTA data from the DR1 release (\citealt{2012A&A...544A.156M}), to select samples of both quasars and LBGs at $z\sim4$. 
This catalog covers an area of $\sim$ 2 $\rm deg^2$ and contains several 
photometric measurements. Specifically in this paper, we use the $u^*$-band $3^{''}$ diameter aperture apparent magnitude measured on the image obtained with MegaCam \citep{2003SPIE.4841...72B} on the Canada-France-Hawaii Telescope (CFHT), and the $3^{''}$ diameter aperture apparent magnitudes of the $g'$-, $r'$-, and $z'$-bands  \citep{2007ApJS..172....9T}
 measured on the image obtained with the Subaru Suprime-Cam \citep{2002PASJ...54..833M}. The 5$\sigma$ limiting AB apparent magnitudes are $u^*$ = 26.5, $g'$ = 26.5, $r'$ = 26.6, and $z'$ = 25.1 ($3^{''}$ diameter aperture). 
We also use the Advanced Camera for Surveys (ACS) catalog \citep{2007ApJS..172..196K, 2007ApJS..172..219L}
 when we select the low-luminosity quasars to separate galaxies from point sources (see \citealt{2011ApJ...728L..25I,2012ApJ...756..160I} for more details). 
\subsection{Selection for Quasars at $z\sim4$}
In order to calculate the bias factor for low-luminosity quasars at $z\sim4$, we need a low-luminosity quasar sample. \cite{2011ApJ...728L..25I} identified eight low-luminosity quasars through spectroscopic follow-up observations for their optical color-selected photometric quasar candidates. Additional low-luminosity quasars were also found by using  the multi-wavelength imaging data including the optical broad, intermediate, narrowband, near-infrared and Spitzer/IRAC photometric measurements \citep{2012ApJ...755..169M}. 
To select  low-luminosity quasars at $z \sim4$ in the COSMOS field we used the following selection criteria:
\begin{equation}
-24.0<M_{\tt 1450}<-22.0,
\end{equation}
and,
\begin{equation}
3.1<z<\rm 4.5.
    \end{equation}
 Note that we adopt both  $z$ and $M_{\tt 1450}$ given in Masters et al. (2012). 
Then we reject objects which lie in masked regions (Capak et al. 2007). As a result, we selected 25 quasars at $3.1<z<4.5$. Among them, 16 objects are spectroscopically confirmed quasars. We refer this sample as the spectroscopic sample while the sample of 25 quasars is as the total one. Total sample have been selected by 
utilizing the 29-band photometric data to remove contaminants. Since they have been selected by utilizing such a lot of photometric data, we consider that the contamination rate for total sample is to be very low. Table 1, Figure 1, and Figure 2 show properties of these quasars, their redshift distribution, and their magnitude distribution, respectively. The median, mean, and the standard deviation of the redshift of the 25 quasars (16 spectroscopically confirmed quasar redshift) are 3.45 (3.59), 3.59 (3.68), and 0.40 (0.40), respectively. 
\begin{table}[!hbt]
\begin{center}
\renewcommand{\tabcolsep}{1pt}
\caption{Properties of the low-luminosity quasars at $z\sim4$}
\begin{tabular}{ccc@{\hspace{0.5cm}}c@{\hspace{0.5cm}}c@{\hspace{0.5cm}}c@{\hspace{0.5cm}}c@{\hspace{0.5cm}}c@{\hspace{0.1cm}}c@{\hspace{0.1cm}}c@{\hspace{0.01cm}}c} \hline\hline
       ID$^{a}$ &  & &R.A.&  Decl. & $M_{\tt 1450}$ &  $z_{\rm sp}$&$z_{\rm adopt}^{b}$ & \\
        &&& (deg) &(deg) & (mag)& &  \\ \hline
298002 &&& 150.43706 & 1.649305 &$-22.34$ &3.89 &3.89& \\
329051 &&& 150.16891 & 1.774590 &$-22.71$ &-- &4.35& \\
330806 &&& 150.10738 & 1.759201 &$-22.85$ &4.14 &4.14& \\
381470 &&& 149.85396 & 1.753672 &$-22.11$ &-- &3.30& \\
422327 &&& 149.70151 & 1.638375 &$-22.49$ &3.20 &3.20& \\
507779 &&& 150.48563 & 1.871927 &$-23.78$ &4.45 &4.45& \\
519634 &&& 150.27715 & 1.958373 &$-22.61$ &-- &3.40& \\
710344 &&& 150.62828 & 2.006204 &$-22.06$ &-- &3.45& \\
804307 &&& 150.00438 & 2.038898 &$-23.56$ &3.50 &3.50& \\
887716 &&& 149.49590 & 1.968019 &$-22.38$ &-- &3.23& \\
1046585 &&& 149.85153 & 2.276400 &$-22.39$ &3.37 &3.37& \\
1060679 &&& 149.73622 & 2.179933 &$-22.23$ &4.20 &4.20& \\
1110682 &&& 149.50595 & 2.185332 &$-22.71$ &-- &3.28& \\
1159815 &&& 150.63844 & 2.391350 &$-22.98$ &3.65 &3.65& \\
1163086 &&& 150.70377 & 2.370019 &$-23.00$ &3.75 &3.75& \\
1271385 &&& 149.86966 & 2.294046 &$-23.42$ &3.35 &3.35& \\
1273346 &&& 149.77692 & 2.444306 &$-22.65$ &4.16 &4.16& \\
1371806 &&& 150.59184 & 2.619375 &$-22.10$ &-- &3.12& \\
1465836 &&& 150.13036 & 2.466012 &$-22.56$ &3.86 &3.86& \\
1575750 &&& 150.73715 & 2.722578 &$-22.78$ &3.32 &3.32& \\
1605275 &&& 150.62006 & 2.671402 &$-22.82$ &3.14 &3.14& \\
1657280 &&& 150.24078 & 2.659058 &$-22.74$ &3.36 &3.36& \\
1719143 &&& 149.75539 & 2.738555 &$-22.26$ &3.52 &3.52& \\
1730531 &&& 149.84322 & 2.659095 &$-22.15$ &-- &3.51& \\
1743444 &&& 149.66605 & 2.740230 &$-22.54$ &-- &3.15& \\
\hline
              \end{tabular}

      \end{center}
         $^{a}$ ID for Table 3 of \cite{2012ApJ...755..169M}.\\
         $^{b}$ Redshift given by Masters et al. (2012) and adopted in our analysis.
\end{table}

\subsection{Selection for Lyman Break Galaxies at $z\sim4$}
We also select a  sample of Lyman break galaxies (LBGs) in the redshift range $3.1<z<4.5$ in the COSMOS field, utilizing the two color diagram of $r'-z'$ vs. $g'-r'$ (Figure 3). For the selection of LBGs at $z\sim4$, we adopt the following selection criteria:
\begin{eqnarray}
149.411400<{\tt R.A. (degree)}<150.826934,\\
1.49878<{\tt Decl. (degree)}<2.91276,\\
      z'<25.0,\\
      u^*\geq 27.05, \\
        g'-r' \geq 1.0, \\  
     g'-r'> 1.1(r'-z')+1.1, 
            \end{eqnarray}  
and,
   \begin{eqnarray}     
              r'-z'\leq 1.5,
         \end{eqnarray}
where $u^{*}$ = 27.05 corresponds to the 3$\sigma$ limiting magnitude in the $u^{*}$-band. The criterion (8) is adopted to select LBGs without significant contamination from low-$z$ elliptical galaxies (see Figure 3). In order to remove
         low-$z$ objects, we add the criteria (6), (7), and (9). These selection criteria are adjusted to select LBGs with photometric redshifts ($z_{\rm ph}$), whose distribution is similar to the redshift distribution of our quasar sample. We also remove eight spectroscopically confirmed quasars satisfying the above criteria from the LBG sample, because those 8 objects are not LBGs apparently. Note that seven objects among the eight removed objects are included in our quasar sample while the remaining one object is not, because its magnitude ($M_{1450}=-20.91$) is out of the magnitude range of our quasar sample ($-22<M_{\rm 1450}<-24$).

  As a result, we obtain a sample of  835 LBGs at $z\sim4$ in the COSMOS field. We use the upgraded photometric redshift (originally described in Ilbert et al. 2009 with including the new UltraVISTA DR1 data) to investigate the photometric redshift distribution of the color-selected galaxies. Figure 4 shows the photometric redshift distribution of color-selected galaxies. The median, mean, and the standard deviation of the redshift of the color-selected galaxies are 3.59, 3.13, and 1.32, respectively. As shown in Figure 4, there are some low-$z$ ($z<1$) galaxies in the color-selected LBG sample. We consider that most of these contaminants are low-$z$ elliptical galaxies, being inferred from the color track of the model elliptical galaxy shown in Figure 3. Note that it is difficult to remove these low-$z$ contaminants by modifying the adopted color-selection criteria, because the optical colors of LBGs and those contaminants are similar. The median, mean, and the standard deviation of the photometric redshift of the color-selected galaxies, after removing objects whose photometric redshifts are below 3, are 3.67, 3.71, and 0.62, respectively. These results are similar to that of low-luminosity quasars in this work. The typical error of the photometric redshift for the color-selected LBGs is $\sim0.1$. There are 67 spectroscopically confirmed objects in our color-selected galaxies and their redshift distribution is shown in Figure 5. We confirm that the spectroscopic redshift distribution of color-selected galaxies is also similar with that of low-luminosity quasars in this paper.
  
\begin{figure}[!t]
\begin{center}
\includegraphics[bb= 35 0 570 388,clip,width=9.4cm]{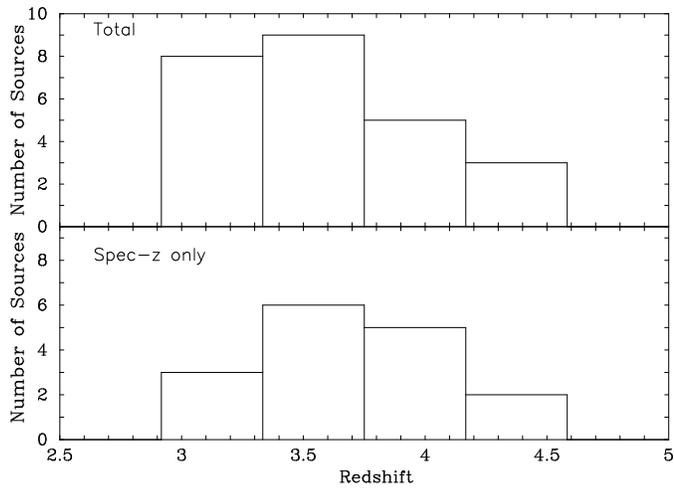}
\caption{Redshift distribution of the low-luminosity quasars at $3.1<z<4.5$ used in this work. Upper and lower panels show the redshift distribution of  25 low-luminosity quasars (total) and 16 spectroscopically confirmed low-luminosity quasars, respectively.}  
\end{center}
\end{figure}
  
    Figure 6 shows the magnitude distribution of the color-selected galaxies used in this paper and we also confirm that the magnitude distribution of the spectroscopic sample
does not drops at a much brighter limit than that of the full photometric
sample. These results may be somewhat surprising, in the sense that
   any spectroscopic sample tends to be brighter than the 
   photometric sample in the same survey generally. However in
   our case, we are now focusing only on relatively bright LBGs 
   even for the photometric sample, 
   whose magnitude is much brighter than the limiting magnitude 
   of the COSMOS survey. This is because we would like to remove 
   most contaminants from our sample and select our sample 
   whose errors of photometric redshift are as small as possible. 
   In addition, it is also expected that bright LBGs and the 
   quasar sample show a strong correlation because it is reported 
   that the LBG clustering becomes strong with increasing UV luminosity (e.g., Ouchi et al. 2004), 
   and then it is considered that brighter LBGs 
exist in more massive dark matter halos. 
   Therefore bright LBGs are useful to investigate the clustering 
   of quasars and galaxies.
 As our measured CCF and ACF will become weaker than the real CCF and ACF due to these low-$z$ galaxies, we correct our CCF and ACF by utilizing the contamination rate for color-selected galaxies (see Sections 3.1 and 3.2).
\begin{figure}[!t]
\begin{center}
\includegraphics[bb= 60 40 430 299,clip,width=9.7cm]{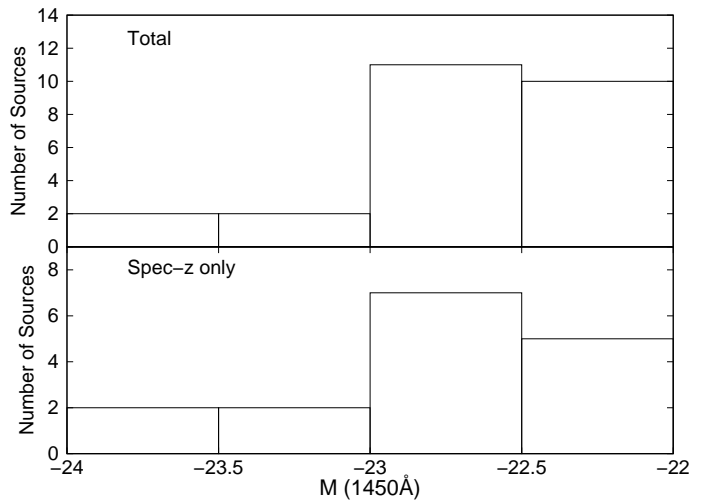}
\caption{Magnitude distribution of the low-luminosity quasars at $3.1<z<4.5$ used in this work. Upper and lower panels show the magnitude distribution of  25 low-luminosity quasars (total) and 16 spectroscopically confirmed low-luminosity quasars, respectively.}  
\end{center}
\end{figure}

\section{Clustering Analysis of quasars and LBGs }
\subsection{Quasar-LBG Two-Point Angular Cross-Correlation Function at $z\sim4$}
Using the samples of quasars and LBGs described in Sections 2.2 and 2.3, we calculate the quasar-LBG two-point angular CCF, $\omega_{\rm QL}$ ($\theta$), at $z\sim4$ using the following equation (\citealt{1999MNRAS.305..547C,2008ApJ...673L..13F}):
\begin{figure}[!htb]
\begin{center}
\includegraphics[bb=30 0 400 330,clip,width=10cm]{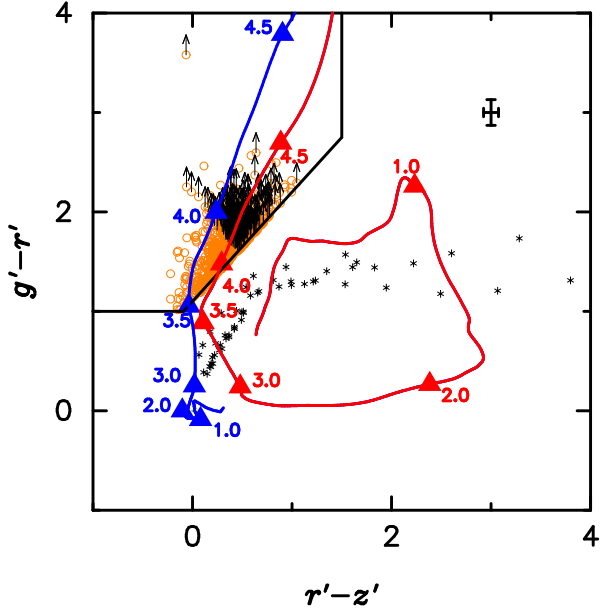}

\caption{Two-color diagram of $g'-r'$ vs. $r'-z'$, that we use for $z\sim4$ LBG selection. Orange circles denote the LBG candidates with $z'<25.0$. Orange circles with a black arrow show the LBG candidates which are not detected in $g'$-band at 3 sigma limiting magnitude, $g'=27.05$ (242 objects among 835 objects). For those cases, the 3$\sigma$ lower limit of their $g'-r'$ color is plotted. The  blue and red lines are the color track of the model star-forming galaxy (where the instantaneous-burst model of \citealt{2003MNRAS.344.1000B} with a metallicity of $Z = 0.02$, an age of 0.025 Gyr and a Ly$\alpha$ equivalent width $=21\rm \AA$ (\citealt{2012ApJ...751...51J}) are adopted) and elliptical galaxy colors (where the stellar population model of \citealt{2003MNRAS.344.1000B} with a metallicity of $Z = 0.02$ and an exponential decay time of $\tau$= 1 Gyr are adopted, and the ages of the model elliptical is 8 Gyr). The IGM absorption is corrected by adopting the model of \cite{1995ApJ...441...18M}, for both color tracks. Triangles denote the color of the model elliptical and star-forming galaxy at $z=1.0$, 2.0, 3.0, 3.5, 4.0, and 4.5, respectively. Black asterisks show colors of G, K, and M-type stars (\citealt{1998PASP..110..863P}). The black solid line shows our photometric criteria used to select LBG candidates at $z\sim4$. The error bar of the upper right side in this Figure denotes 1$\sigma$ error for the $g'-r'$ and $r'-z'$ of our LBG candidates.}

\end{center}
\end{figure}

\begin{equation}
\rm \omega_{\rm QL} (\theta) = \it\frac{ \langle D_{\rm Q}D_{\rm L}\rangle }{\it\langle D_{\rm Q}R  \rangle}-\rm1,
\end{equation}
where $ \langle D_{\rm Q}D_{\rm L}\rangle$ and $ \langle D_{\rm Q}R\rangle$ are the normalized quasar-LBG and quasar-random number of pairs defined
as follows:
\begin{equation}
 \langle D_{\rm Q}D_{\rm L}\rangle = \frac{ D_{\rm Q}D_{\rm L}(\theta)}{\it N_{\rm Q}\it N_{\rm L}},
 \end{equation}
 and,
 \begin{equation}
  \langle D_{\rm Q}R\rangle = \frac{ D_{\rm Q}R(\theta)}{\it N_{\rm Q}\it N_{\rm R}},
\end{equation} 
where $ D_{\rm Q}D_{\rm L} (\theta)$ and $ D_{\rm Q}R (\theta)$ are the number of data-data and data-random pairs at angular separation $\theta\pm \Delta\theta$, respectively. In the equations (11) and (12), $ N_{\rm Q}$, $ N_{\rm L}$, and $ N_{\rm R}$ are the total number of the quasar, LBG, and random sample, respectively. We create the 100,000 random samples which are avoiding the masked regions and we calculate the quasar-LBG CCF.  The errors in $\omega_{\rm \small QL} (\theta)$ are estimated by the bootstrap method as follows (\citealt{1986MNRAS.223P..21L}):
\begin{equation}
\sigma_{\omega_{\rm QL}}=\Bigl \{\sum_{i=1}^{N}\frac{[\omega_{i} (\theta)-\langle\omega (\theta)\rangle]^{2}}{N-1}\Bigl \}^{1/2},
\end{equation}
where $N$ is the number of bootstrap samples and $\langle\omega (\theta)\rangle$ is calculated as follows:
\begin{equation}
\langle\omega (\theta)\rangle=\sum_{i=1}^{N}\frac{\omega_{i}(\theta)}{N}.
\end{equation}
We use $N=1,000$, and Figure 5 shows the result of the quasar-LBG two-point angular CCF at $z\sim4$. 
Previous investigators mentioned that the Poisson error becomes 
increasingly inaccurate at larger scale (e.g., Mountrichas et al. 2009).
We also confirmed that the errors which are calculated by the bootstrap method
are about two times larger than the Poisson errors.
We calculate the CCF for both total and spectroscopic samples.
We summarize $ \omega_{\rm QL} (\theta)$ and errors of $ \omega_{\rm QL} (\theta)$ in Table 2.
The observed two-point angular CCF is approximated by a power law form at large scales as follows:
\begin{equation}
\omega_{\rm QL} (\theta) = A_{\omega}^{\rm CCF}(\theta^{-\beta}-\rm C),
\end{equation}
where $\beta$ is fixed to be 0.8 (Francke et al. 2008) and C is the integral constraint (\citealt{1977ApJ...217..385G}). To avoid the one-halo term and negative value at the second bin, we do not use $\rm \omega_{\rm QL} (\theta)$ on smaller scales ($\rm angular \ separation<80$ arcsec) to calculate $A_{\omega}^{\rm CCF}$. Here we estimate the integral constraint as follows:
  \begin{figure}[!htb]
\begin{center}
\includegraphics[bb= 20 300 540 728,clip,width=8.9cm]{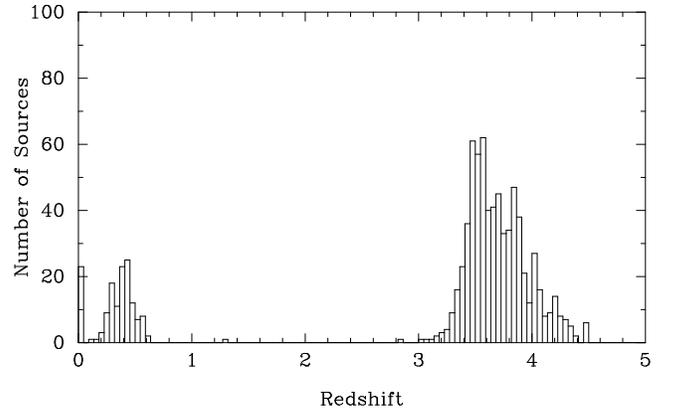}
\vspace{-2cm}
\caption{Photometric redshift distribution of the 835 color-selected galaxies used in this paper.}
\end{center}
\end{figure}
  \begin{figure}[!htb]
\begin{center}
\includegraphics[bb= 20 0 540 368,clip,width=8.8cm]{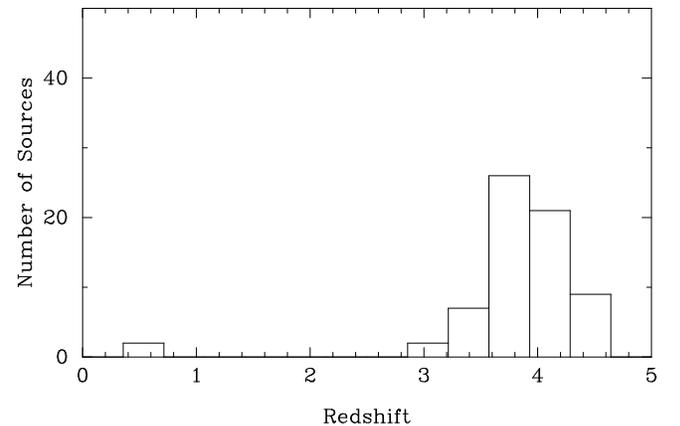}

\caption{Spectroscopic redshift distribution of the 67 objects among the 835 color-selected galaxies used in this paper.}
\end{center}
\end{figure}

\begin{equation}
 \rm C= \frac{\Sigma \it RR (\theta) \theta^{\rm -0.8}}{\Sigma\it RR (\theta)}.
\end{equation}
We calculate C and a value of C = 0.00601 in this case.
Using equation (15) and (16), we calculate $A_{\omega}^{\rm CCF}$ to fit the two-point angular CCF. As a result, we obtain $A_{\omega}^{\rm CCF}$ =  1.77$^{+1.66}_{-0.86}$ and 3.33$^{+3.06}_{-1.60}$ for the total and the spectroscopic samples, respectively.
\begin{figure}[h]
\begin{center}
\includegraphics[bb= 60 40 430 299,clip,width=9.7cm]{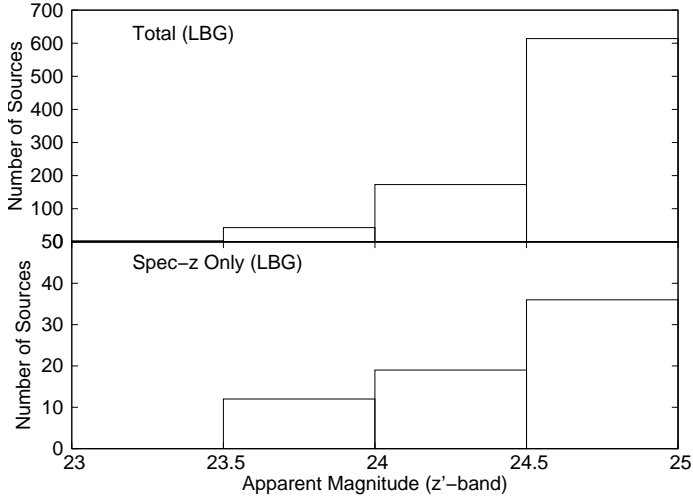}
\caption{Magnitude distribution of the color-selected galaxies used in this paper. Upper and lower panels show the magnitude distribution of  835 color-selected galaxies (total) and 67 spectroscopically confirmed galaxies at $z\sim4$, respectively.}  
\end{center}
\end{figure}

 Since we use the photometric sample of LBGs, the derived correlation
amplitude is affected by some contamination of galaxies at lower redshifts.
Accordingly the effect of the contamination on the derived correlation
amplitude should be corrected. 
We estimated the contamination rate, $f_{\rm c}$, utilizing the distribution
of spectroscopic redshifts in our LBG sample (see
Figures 5). In a naive estimate, the contamination rate is calculated
through the following formula using the spectroscopic redshift of our
color-selected LBGs:
$f_{\rm c} = (N(z_{\rm sp} < 3.1) + N(z_{\rm sp} > 4.5))/N_{\rm total}$,
where the numbers of contaminating objects are $N(z_{\rm sp} < 3.1) =
2$ and $N(z_{\rm ph} > 4.5) = 0$, among our color-selected and spectroscopically confirmed objects
($N_{\rm total} = 67$). Therefore the contamination rate in our color-selected LBG sample is estimated to be
$f_{\rm c} = 2/67 \sim0.03$. In order to calculate $A_{\omega}$ (CCF) accurately, 
we need to calculate the contamination rate in the total sample for low-luminosity quasars. 
Therefore we calculate the contamination rate for the total sample, $f_{\rm cq}$, as follows:
\begin{equation}
f_{\rm cq}= \int^{M=-22}_{M=-24}\frac{n_{\rm total} (M)[f_{\rm r}\Phi_{\rm LBG} (M)/\Phi_{\rm QSO}(M)]dM}{N_{\rm total}},
\end{equation}
where $f_{\rm r}$, $n_{\rm total}(M)$, $\Phi_{\rm LBG}(M)$, $\Phi_{\rm QSO}(M)$, and $N_{\rm total}$ are the fraction which high-redshift galaxies pass quasar selections, the magnitude distribution of low-luminosity quasars for total sample, the luminosity function of LBGs, the luminosity function of quasars, and the total number of low-luminosity quasars, respectively. Masters et al. (2012) estimated $f_{\rm r}$ and they found that only 2 objects among 386 spectroscopically confirmed high-redshift galaxies are pass their quasar selection. Therefore $f_{\rm r}$ is estimated to be $\sim0.005$. We use $\Phi_{\rm LBG}(M)$ which are derived by \cite{2007ApJ...670..928B} and $\Phi_{\rm QSO}(M)$  which are derived by Masters et al. (2012), in this paper. Using above results and Equation (17), we calculate $f_{\rm cq}$ and the calculated $f_{\rm cq}$ is $\sim0.05$. Since the contamination rate for the total sample is so low, 
we do not worry about impact on the CCF due to include high-redshift 
galaxies in the total sample. Even if some high-redshift galaxies 
such as bright LBGs are including in the total sample, we consider 
that the clustering signal will not become weak by the contaminants such as 
bright LBGs because the LBG ACF becomes strong with increasing UV luminosity. 
From the above, we do not use the contamination rate for the total sample 
though we use the contamination rate for LBGs to calculate the $A_{\omega}$ (CCF).

For taking the estimated contamination rate into account, we calculate $A_{\omega}$ (CCF) using the following equation,
\begin{equation}
A_{\omega} (\rm CCF)=\it\frac{A_{\omega}^{\rm CCF}}{(\rm1-\it f_{\rm c})}.
\end{equation}
We then obtain $A_{\omega} (\rm CCF)=$1.82$^{+1.71}_{-0.88}$ and 3.43$^{+3.16}_{-1.64}$ for the total and the spectroscopic samples, respectively.
The results of $A^{\rm CCF}_{\omega}$ and $A_{\omega} (\rm CCF)$ are given in Table 3. We find that the observed CCF, $\omega_{\rm QL} (\theta)$ is larger than 0 at smaller scales. However, there is some possibility that this result is only caused by the position of LBGs because of the small number of quasars (i.e., this result is not caused by the position of quasars). 

In order to confirm that this result ($\omega_{\rm QL} (\theta)>0$) is caused by the position of quasars and LBGs, we also generate 16 random points for quasars ($R_{\rm Q}$) and calculate the random-LBG CCF 6000 times. Using $ \rm \omega_{\rm R_{\rm Q}L} (\theta) = \it\frac{ \langle R_{\rm Q}D_{\rm L}\rangle }{\it\langle R_{\rm Q}R  \rangle}-\rm1$, we find that $\omega_{\rm R_{\rm Q}L} (\theta)\sim0$ at all scale while the errors are large at smaller scales (see Figure 7). We also calculate the probability of $A_{\omega}$ (CCF) for random quasars
is larger than that of $A_{\omega}$ (CCF) for real quasars.
As a result, this probability is $\sim20\%$. Since this probability seems to be high, we treat $A_{\omega} (\rm CCF)+\sigma_{A_{\omega} (\rm CCF)}$ as the $86\%$ upper limit. We then calculate the $86\%$ upper limits of the spatial correlation length and the bias factor for low-luminosity quasars in the same way. 

  \begin{figure*}[]
\begin{center}
\includegraphics[bb= 50 0 550 408,clip,width=12.5cm]{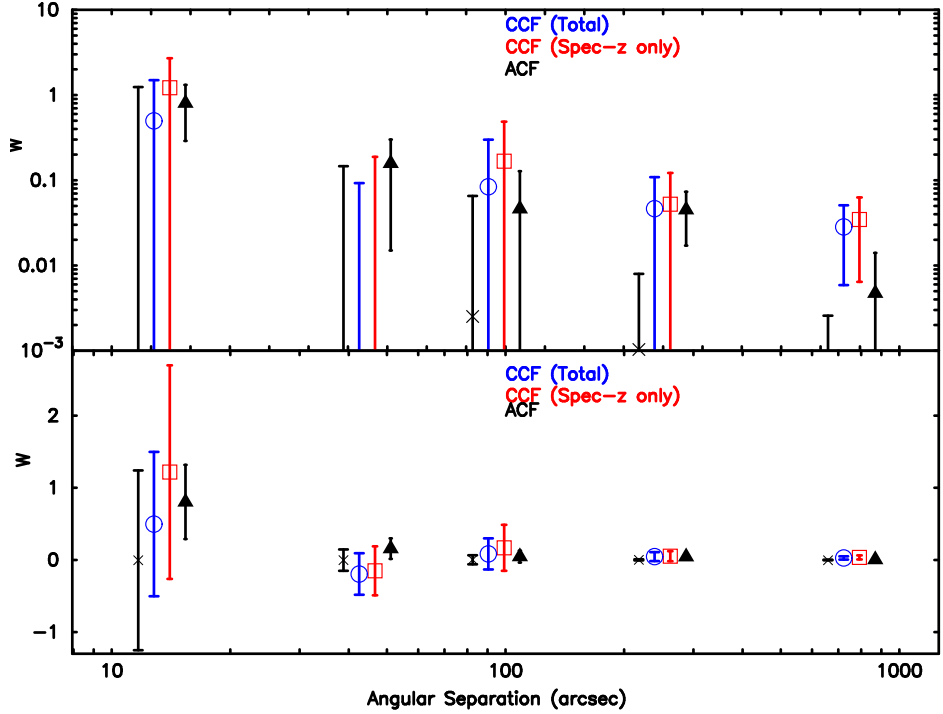}
\caption{Two-point angular quasar-LBG CCF (red open squares and blue open circles), the random-LBG CCF (black crosses), and the LBG ACF (filled black triangles) at $z\sim4$ in the COSMOS field. Top and bottom panel shows the two-point angular quasar-LBG CCF and the LBG ACF in double-logarithmic and semilogarithmic graph, respectively. The data are shown with slight shifts in the horizontal direction for clarity. }
\end{center}
\end{figure*}

        \begin{table*}[!hbt]
\begin{center}
\renewcommand{\tabcolsep}{1pt}
\caption{Summary of  the $\omega_{\rm QL}$ at $z\sim4$ }
\begin{tabular}{ccc@{\hspace{0.5cm}}c@{\hspace{0.5cm}}c@{\hspace{0.5cm}}c@{\hspace{0.5cm}}c@{\hspace{0.5cm}}c@{\hspace{0.5cm}}c@{\hspace{0.5cm}}c} \hline \hline
       $\theta$  &&& $\omega_{\rm QL}$ &$\sigma_{\omega_{\rm QL}}$ &$ D_{\rm Q}D_{\rm L}(\theta)$& $\omega_{\rm QL}$&$\sigma_{\omega_{\rm QL}}$ &$ D_{\rm Q}D_{\rm L}(\theta)$&  \\ 
      (arcsec) &  & &(Total)&(Total)   &(Total)&($z_{\rm sp}$ only)  &($z_{\rm sp}$ only)  &($z_{\rm sp}$ only) & \\\hline
 14  &&& 0.4970 &0.9997&2&1.2178& 1.4805 & 2& \\
  46 &&&  $-0.1950$ &0.2877&9& $-0.1506$ & 0.3389&6 & \\
   99 &&& 0.0838 &0.2156& 32&0.1674 & 0.3189 &22 &\\
   262 &&& 0.0463 &0.0625&382&0.0523 & 0.0693& 242& \\
 792 &&&  0.0284 &0.0225& 2803&0.0347 & 0.0282 &1832&\\ \hline
              \end{tabular}

       \end{center}
       
\end{table*}

\begin{table*}[!hbt]

\begin{center}
\caption{Summary of the quasar-lbg ccf }
\scalebox{0.9}[0.9]{
\begin{tabular}{ccc@{\hspace{0.2cm}}c@{\hspace{0.2cm}}c@{\hspace{0.2cm}}c@{\hspace{0.2cm}}c@{\hspace{0.2cm}}c@{\hspace{0.2cm}}c@{\hspace{0.2cm}}c} \hline \hline
         $ N_{\rm Q}$&$ N_{\rm LBG}$ &    QSO Magnitude &LBG Magnitude& Redshift & $A^{\rm CCF}_{w}$  & $A_{w}$ & $A_{w}$ &$r_{0}$ & \\
        &&  &&& &&($86\%$ upper limit)&($h^{-1}$ Mpc) &  \\ \hline
           16$^{a}$& 835&   $-24.0<M_{\rm 1450}<-22.0$ & $z'<25.0$&$3.1<z<4.5$ &3.33$^{+3.06}_{-1.60}$ &3.43$^{+3.16}_{-1.64}$ & $<6.59$&$<10.72$&\\
           25$^{b}$& 835&   $-24.0<M_{\rm 1450}<-22.0$ & $z'<25.0$&$3.1<z<4.5$ &1.77$^{+1.66}_{-0.86}$ &1.82$^{+1.71}_{-0.88}$ & $<3.53$&$<7.60$&\\
           \hline \\
                        \end{tabular}
}
\end{center}
$^{a}$Number of the spectroscopically confirmed quasars.\\
$^{b}$Total number of the quasars.\\
\end{table*}
\begin{table}[!hbt]
\begin{center}
\renewcommand{\tabcolsep}{1pt}
\caption{Summary of  the  $\omega_{\rm LL}$ at $z\sim4$ }
\begin{tabular}{ccc@{\hspace{0.5cm}}c@{\hspace{0.5cm}}c@{\hspace{0.5cm}}c@{\hspace{0.5cm}}c} \hline \hline
       $\theta$  &&& $\omega_{\rm LL}$ &$\sigma_{\omega_{\rm LL}}$  &$ D_{\rm L}D_{\rm L} (\theta)$   \\ 
      (arcsec) &  & &&   &  &   \\\hline
 14  &&& 0.8034 &0.5134& 40\\
  46 &&&  0.1576 &0.1426& 220\\
   99 &&& 0.0463 &0.0816& 512\\
   262 &&& 0.0452 &0.0280& 5851\\
 792 &&&  0.0047 &0.0093& 42091\\ \hline
              \end{tabular}

      \end{center}
\end{table}

\subsection{LBG Two-Point Angular Auto-Correlation Function at $z\sim4$}
 To constrain the quasar triggering mechanism, we have to calculate the CCF and the LBG two-point angular auto-correlation function (ACF). Since we have calculated the CCF in Section 3.1, we calculate the LBG ACF at $z\sim4$ in Section 3.2.
In order to calculate the LBG two-point angular ACF at $z\sim4$,
we use the \cite{1993ApJ...412...64L} estimator:

\begin{equation}
\rm \omega_{\rm LL} (\theta) = \frac{ \it \langle DD  \rangle -\rm2\langle\it DR  \rangle + \langle RR  \rangle }{\it \langle RR  \rangle},
\end{equation}
where $ \langle DD \rangle$, $ \langle DR \rangle$, and $ \langle RR  \rangle$ are the normalized data-data, data-random, and random-random number of pairs. Those are defined
as follows:
\begin{eqnarray}
 \langle DD \rangle = \frac{ D_{\rm L}D_{\rm L} (\theta)}{ \it N_{\rm L}(\it N_{\rm L}\rm-1)/2},\\
 \langle DR \rangle = \frac{ D_{L}R (\theta)}{\it N_{\rm L}\it N_{\rm R}},
 \end{eqnarray}
 and,
 \begin{equation}
  \langle RR  \rangle = \frac{ RR (\theta)}{\it N_{\rm R}(\it N_{\rm R}-\rm 1)/2},
\end{equation}
where $ D_{L}D_{L} (\theta)$, $ D_{L}R (\theta)$, and $ RR (\theta)$ are number of data-data, data-random, and random-random pairs at angular separation $\theta\pm \Delta\theta$, respectively. In equations (20)--(22), $N_{\rm L}$ and $ N_{\rm R}$ are the total number of data in the LBG and random sample, respectively. 
Using 100,000 random samples which are avoiding the masked regions, we calculate the two-point ACF of LBGs. The obtained results are shown in Figure 7 and Table 4. We find that the quasar-LBG two-point CCF is similar with the LBG two-point ACF at $z\sim4$. The observed two-point angular ACF is also approximated by a power law form at large scales as follows:
\begin{equation}
\omega_{\rm LL} (\theta) = A_{\omega}^{\rm ACF}(\theta^{-\beta}-\rm C),
\end{equation}
where $\beta$ is fixed to be 0.8 (Francke et al. 2008) and C (which gives a value of C = 0.00601 in this case) is the integral constraint. To avoid the one-halo term, we do not use $\rm \omega_{\rm LL} (\theta)$ on small scales ($\rm angular \ separation<40$ arcsec) to calculate $A_{\omega}^{\rm ACF}$. We calculate the correlation amplitude of the LBG ACF, $A_{\omega}^{\rm ACF}$ using equations (16) and (23). As a result, $A_{\omega}^{\rm ACF}$ is $3.65^{+2.23}_{-1.38}$. In addition, we calculate the correlation amplitude for ACF, $A_{\omega}$ (ACF) as follows:
\begin{equation}
A_{\omega} (\rm ACF)=\it\frac{A_{\omega}^{\rm ACF}}{(\rm1-\it f_{\rm c})^{\rm2}},
\end{equation}
the calculated $A_{\omega} (\rm ACF)$ is 3.88$^{+2.37}_{-1.47}$. These results are listed in Table 5.

\begin{table*}[!hbt]

\begin{center}
\caption{Summary of the LBG ACF}

\begin{tabular}{ccc@{\hspace{0.2cm}}c@{\hspace{0.2cm}}c@{\hspace{0.2cm}}c@{\hspace{0.2cm}}c@{\hspace{0.2cm}}c@{\hspace{0.2cm}}c} \hline \hline
         $ N_{\rm LBG}$ &   LBG Magnitude& Redshift &$A^{\rm ACF}_{w}$ & $A_{w}$ &$r_{0}$ &  \\
        &&  &&& ($h^{-1}$ Mpc) &  \\ \hline
 835&   $z'<25.0$ &$3.1<z<4.5$ & $3.65^{+2.23}_{-1.38}$ & $3.88^{+2.37}_{-1.47}$ & $6.52^{+3.16}_{-1.96}$&\\
           \hline \\
                        \end{tabular}

\end{center}

\end{table*}
\begin{table*}[!hbt]
\begin{center}
\caption{Summary of the bias factor }
\begin{tabular}{ccc@{\hspace{0.2cm}}c@{\hspace{0.2cm}}c@{\hspace{0.2cm}}c@{\hspace{0.2cm}}c@{\hspace{0.2cm}}c@{\hspace{0.2cm}}c@{\hspace{0.2cm}}c} \hline \hline
         $ N_{\rm Q}$&$ N_{\rm LBG}$ &  &  QSO Magnitude &LBG Magnitude& Redshift & $b_{\rm QSO}$ & $b_{\rm LBG}$  &$b_{\rm QL}$\\
        \hline
           16& 835&&  $-24.0<M_{\rm 1450}<-22.0$ & $z'<25.0$&$3.1<z<4.5$ &$<10.50$& $4.92^{+2.07}_{-1.29}$ &$<7.69$\\
           25& 835&&  $-24.0<M_{\rm 1450}<-22.0$ & $z'<25.0$&$3.1<z<4.5$ &$<5.63$& $4.92^{+2.07}_{-1.29}$ &$<5.65$\\
           \hline \\
                        \end{tabular}
                        \end{center}
                        \end{table*}
\subsection{The Spatial Correlation Function}
In order to calculate the bias factor for LBGs and low-luminosity quasars, we calculate the spatial correlation function for the ACF and CCF.
The spatial correlation function is given as follows:
\begin{equation}
\xi(r)=(r/r_{\rm 0})^{-\gamma},
\end{equation}
where $r_{\rm 0}$ is the spatial correlation length and $\gamma$ = $\beta$+1. The spatial
correlation function for the ACF is calculated with the following relation (\citealt{1969PASJ...21..221T}),
\begin{equation}
A_{w}(\rm ACF)=\it\frac{H_{\gamma} r_{\rm 0}^{\gamma} \int^{}_{} F(z) r^{\rm1-\it\gamma}_{c}(z)N_{\rm g}^{\rm2}\it(z)E(z)dz}{(c/
H_{\rm 0})[\int^{}_{}N_{\rm g}(z)dz]^{\rm2}},
\end{equation} 
where $r_{c}$ is the comoving radial distance, $N_{\rm g}(z)$ is the redshift distribution of LBGs, and $F(z)$ is the evolution of clustering with redshift, which is assumed to be negligible and is set equal to 1 in this paper. In equation (26), $H_{\gamma}$ and $E(z)$ are calculated as follows:
\begin{equation}
H_{\gamma}=\Gamma(1/2)\frac{\Gamma[(\gamma-1)/2]}{\Gamma(\gamma/2)}, 
\end{equation}
and,
\begin{equation}
E(z)=[\Omega_{m}(1+z)^{3}+\Omega_{\Lambda}]^{1/2}.
\end{equation}
The spatial
correlation function for the CCF is also calculated following the relation (\citealt{1999MNRAS.303..411C});
\begin{equation}
A_{w}(\rm CCF)=\it \frac{H_{\gamma} r_{\rm 0}^{\gamma} \int^{}_{} F(z) r^{\rm1-\it \gamma}_{c}(z)N_{\rm g}\it(z)N_{\rm q}\it(z)E(z)dz}{(c/H_{\rm 0})[\int^{}_{}N_{\rm g}\it(z)dz\int^{}_{}N_{\rm q}\it(z)dz]},
\end{equation}
where $N_{\rm q}(z)$ is the redshift distributions of quasars.
We calculate the spatial correlation lengths of the LBG ACF  and the quasar-LBG CCF using these equations.
As a result, the spatial correlation length for the LBG ACF is  6.52$^{+3.16}_{-1.96}$ $h^{-1}$ Mpc. The $86\%$ upper limits of the spatial correlation lengths for the quasar-LBG CCF are  $7.60$ $h^{-1}$ Mpc and $10.72$ $h^{-1}$ Mpc for the total and the spectroscopic sample, respectively.
These results are also summarized in Tables 3 and 5.

\subsection{The Bias Factor}
We now investigate the bias factor for low-luminosity quasars at $z\sim4$ in the COSMOS field to study the luminosity dependence of the quasar clustering and constrain the triggering mechanism of the quasar activity. In order to investigate the quasar bias factor,
we have to estimate the galaxy bias at the same redshift. In this paper, we estimate the galaxy bias factor at the same redshift using LBGs. 
The bias factors for LBGs and quasars are defined as follows:
\begin{equation}
b_{\rm LBG} (z) = \sqrt[]{\frac{\xi_{\rm L} (8, z)}{\xi_{\rm DM} (8,z)}}, b_{\rm QSO} (z) = \sqrt[]{\frac{\xi_{\rm Q} (8, z)}{\xi_{\rm DM} (8,z)}},\end{equation}
where $\xi_{\rm L}(8,z)$ = $(r_{\rm 0}(z)/8)^{\gamma}$, $\xi_{\rm Q}(8,z)$ = $(r_{\rm 0}(z)/8)^{\gamma}$, and $\xi_{\rm DM}(8,z)$ are the spatial correlation functions of LBGs, quasars, and dark matter halos evaluated at 8 $h^{-1}$ Mpc, respectively.
The correlation function of dark matter halos is as follows (\citealt{1980lssu.book.....P}):
\begin{equation}
\xi_{\rm DM} (8,z) = \frac{\sigma^{2}_{8}(z)}{J_{2}},
\end{equation}
where $J_{2}$ = $72/[(3-\gamma)(4-\gamma)(6-\gamma)2^{\gamma}]$ and $\sigma^2_{8}(z)$ is the dark matter density variance in a sphere with a comoving radius of 8 $h^{-1}$ Mpc. We calculate $\sigma_{8}(z)$ using the following equation:
\begin{equation}
\sigma_{8}(z)=\sigma_{8}\frac{D(z)}{D(0)},
\end{equation}
where $D(z)$ is the linear growth factor scaled to unity at the present time and we calculate it as follows (\citealt{2006ApJ...638..622M}):
\begin{equation}
D(z)=\frac{g_{z}}{g_{\rm0}}\frac{1}{(1+z)}.
\end{equation}
We calculate $g_{z}$ by using the following equation (\citealt{1992ARA&A..30..499C,2006ApJ...638..622M}):
\begin{equation}
g_{z}=\frac{5\Omega_{mz}}{2}\Bigl [\Omega^{4/7}_{mz}-\Omega_{\Lambda z}+(1+\frac{\Omega_{mz}}{2})(1+\frac{\Omega_{\Lambda z}}{70})\Bigl ]^{-1},
\end{equation}
where we also calculate $\Omega_{mz}$ and $\Omega_{\Lambda z}$ as follows (\citealt{2006ApJ...638..622M}):
\begin{equation}
\Omega_{mz}=\Bigl(\frac{H_{\rm0}}{H_{z}}\Bigl)^{2}\Omega_{m}(1+z)^{3}, \Omega_{\Lambda z}=\Bigl(\frac{H_{\rm 0}}{H_{z}}\Bigl)^{2}\Omega_{\Lambda},
\end{equation}
where $H_{z}$ is expressed as follows (\citealt{2006ApJ...638..622M}):
\begin{equation}
H_{z}=H_{\rm0}[\Omega_{m}(1+z)^{3}+\Omega_{\Lambda}]^{1/2}.
\end{equation}
Using equations (30)--(36), we calculate the correlation function of the dark matter halos evaluated at 8 $h^{-1}$ Mpc and the bias factor for LBGs. As a result, the bias factor for the LBGs is 4.92$^{+2.07}_{-1.29}$. Next we describe how to calculate the bias factor for low-luminosity quasars at $z\sim4$ using the bias factor for LBGs.
In order to calculate the bias factor for low-luminosity quasars, we use the relations as follows (\citealt{2009MNRAS.394.2050M}):
\begin{equation}
b^{2}_{\rm LBG}= \sigma^{2}_{\rm 8, LBG}/\sigma^{2}_{\rm8, DM}, \
b_{\rm QSO}b_{\rm LBG}=b_{\rm QL}^2,
\end{equation}
where $b_{\rm QL}$ is the bias factor for the quasar-LBG CCF and this is calculated as follows:
\begin{eqnarray}
b_{\rm QL}^{2}=\sigma^{2}_{\rm8,QSO-LBG}/\sigma^{2}_{\rm8, DM}.
\end{eqnarray}
 We then obtain the $86\%$ upper limits of  $b_{\rm QL}$  $=5.65$ and $7.69$ for the total and the spectroscopic sample, respectively. Based on these values, we derive 
 the $86\%$ upper limits of $b_{\rm QSO}$ $=5.63$ and $10.50$ for the total and spectroscopic sample, respectively. 
The results of the bias factors for $b_{\rm LBG}$, $b_{\rm QL}$, and $b_{\rm QSO}$ are summarized in Table 6.

\subsection{The Typical Dark Matter Halo Mass}
The inferred bias factor for low-luminosity quasars can be used to calculate the typical dark matter halo mass ($M_{\rm DM}$). To calculate the typical dark matter halo mass in which low-luminosity quasars at $z\sim4$ reside, we use Equation (8) of \cite{2001MNRAS.323....1S} with parameters which
are recalibrated by  \cite{2005ApJ...631...41T} as follows:
\begin{eqnarray}
 \nonumber b (M, z)=1+\frac{1}{\sqrt{a}\delta_{c} }\Bigl [a\nu^{2}\sqrt{a}+b\sqrt{a}(a\nu^{2})^{(1-c)}\\ -\frac{(a\nu^2)^c}{(a\nu^2)^c+b(1-c)(1-c/2)}\Bigl ],
\end{eqnarray}
where $\nu=\frac{\delta_{c} }{\sigma(m)D(z)}$, $a=0.707$, $b=0.35$, $c=0.80$, and $\delta_{c}\sim1.686$ is the critical overdensity. $\sigma(M)$ is the linear theory rms mass fluctuation on the mass scale $M$ and we calculate it using equations (A8), (A9), and (A10) of \cite{2002MNRAS.331...98V}. The dark matter halo mass can be estimated more accurately by utilizing the halo occupation distribution (HOD) models. However the errors in the derived quasar bias are still large and 
therefore we only use Equation (39). We then obtain the $86\%$ upper limits of log $(M_{\rm DM}/(h^{-1}M_{\odot}))=12.7$ and $13.5$, for the total and the spectroscopic sample, respectively.
\begin{figure}[!t]
\begin{center}
\vspace{-2cm}
\includegraphics[bb= 0 20 750 628,clip,width=10.6cm]{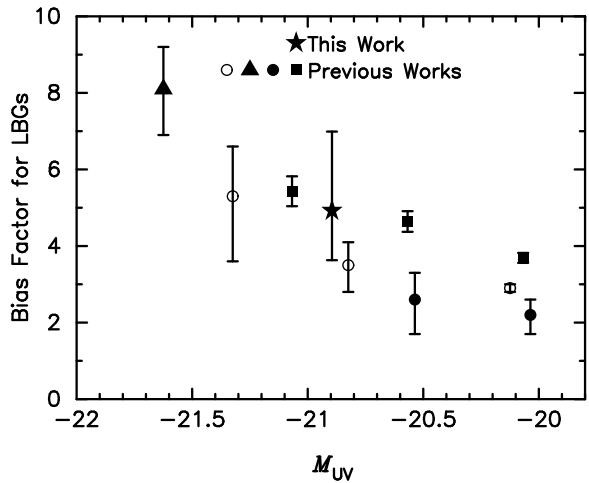}
\caption{Bias factor for LBGs as a function of UV magnitude. Star, square, open circle, triangle, and solid circle points show our result, results of \cite{2009A&A...498..725H},  \cite{2004ApJ...611..685O}, \cite{2005MNRAS.360.1244A}, and \cite{2001ApJ...558L..83O}, respectively.}
\end{center}
\end{figure}

\begin{table*}[!hbt]
\caption{Summary of the reported bias factor for quasars}
\scalebox{0.75}[0.75]{
\begin{tabular}{ccc@{\hspace{0.3cm}}c@{\hspace{0.3cm}}c@{\hspace{0.3cm}}c@{\hspace{0.3cm}}c@{\hspace{0.3cm}}c@{\hspace{0.3cm}}c@{\hspace{0.3cm}}c@{\hspace{0.3cm}}c} \hline \hline
       Sample& &  &  $ N_{\rm Q}$ & Quasar (AGN) Luminosity &  $z$-Range  &Quasar (AGN) Bias  & Type of CF &Reference&\\
        &&&  &(mag or erg s$^{-1}$)&& & && &  \\ \hline
        COSMOS & & & 16 & $-24.0<M_{\rm 1450}<-22.0$ & $3.1<z<4.5$  &$<10.50$& CCF &This work&\\
        COSMOS & & & 25 & $-24.0<M_{\rm 1450}<-22.0$ & $3.1<z<4.5$  &$<5.63$& CCF &This work&\\
        GOODS$^{a}$& & & $25$ & $-30<M_{\rm UV}<-25$ & $1.6<z<3.7$ &$3.9\pm3.0^{b}$& CCF&\cite{2005ApJ...627L...1A}&\\
GOODS& & & $54$ & $-25<M_{\rm UV}<-20$ & $1.6<z<3.7$ &$4.7\pm1.7^{b}$& CCF&\cite{2005ApJ...627L...1A}&\\
2QZ$^{c}$& & & 18,066 & $18.25<b_{J}<20.85$ & $0.3<z<2.2$ &$2.02\pm0.07$& ACF&Croom et al. (2005)&\\
 2QZ& & & $\sim14,000$$^{d}$ & $M_{b_{J}}<-22.5$$^{e}$ & $0.80<z<1.06$ &$1.57^{+0.30}_{-0.37}$& ACF&\cite{2006MNRAS.371.1824P}&\\
        2QZ& & & $\sim14,000$$^{d}$ & $M_{b_{J}}<-22.5$$^{e}$ & $1.06<z<1.30$ &$1.76^{+0.35}_{-0.43}$& ACF&\cite{2006MNRAS.371.1824P}&\\
       2QZ& & & $\sim14,000$$^{d}$ & $M_{b_{J}}<-22.5$ $^{e}$& $1.30<z<1.51$ &$2.13^{+0.29}_{-0.33}$& ACF&\cite{2006MNRAS.371.1824P}&\\
      2QZ& & & $\sim14,000$$^{d}$ & $M_{b_{J}}<-22.5$$^{e}$& $1.51<z<1.70$ &$2.33^{+0.33}_{-0.39}$& ACF&\cite{2006MNRAS.371.1824P}&\\
  2QZ& & & $\sim14,000$$^{d}$ & $M_{b_{J}}<-22.5$ $^{e}$& $1.70<z<1.89$ &$3.02^{+0.45}_{-0.53}$& ACF&\cite{2006MNRAS.371.1824P}&\\
2QZ& & & $\sim14,000$$^{d}$ & $M_{b_{J}}<-22.5$$^{e}$ & $1.89<z<2.10$ &$4.13^{+0.49}_{-0.55}$& ACF&\cite{2006MNRAS.371.1824P}&\\
 SDSS DR1& & & $100,563^{d}$ & $14.5\leq g<21.0$ & $0.4<z<1.0$ &$1.34\pm0.56$& ACF&\cite{2006ApJ...638..622M}&\\
        SDSS DR1& & & $100,563^{d}$ & $14.5\leq g<21.0$ & $1.0<z<1.4$ &$2.20\pm0.26$& ACF&\cite{2006ApJ...638..622M}&\\
        SDSS DR1& & & $100,563^{d}$ & $14.5\leq g<21.0$ & $1.4<z<1.7$ &$2.58\pm0.35$& ACF&\cite{2006ApJ...638..622M}&\\
        SDSS DR1& & & $100,563^{d}$ & $14.5\leq g<21.0$ & $1.7<z<2.1$ &$2.42\pm0.39$& ACF&\cite{2006ApJ...638..622M}&\\
        SDSS DR1& & & $100,563^{d}$ & $14.5\leq g<21.0$ & $2.1<z<2.8$ &$3.12\pm0.80$& ACF&\cite{2006ApJ...638..622M}&\\
        $\rm2QZ+2SLAQ$$^{f}$& & & $22,416$$^{g}$ & $18.25<b_{J}<20.85^{h}$& $0.3<z<2.9$ &$1.5\pm0.2$& ACF&\cite{2008MNRAS.383..565D}&\\
& & & $6,374$$^{i}$ & $20.50<g<21.85^{j}$& & && &&\\
        MUSYC ECDF$-$S$^{k}$ & & & 58 & $R<25.5$ & $2.7<z<3.8$ &$5.5\pm2.0$& CCF&Francke et al. (2008)&\\
        AEIGS$^{l}$& & & 113 & $M_{B}$ $=-20.98$ (median) & $0.7<z<1.4$ &$1.48\pm0.12$& CCF&\cite{2009ApJ...701.1484C}&\\
        XMM$-$COSMOS$^{m}$& & & 538 & $i'<23.0$ & $0.2<z<3.0$ &$2.0\pm0.2$& ACF&Gilli et al. (2009)&\\
        2SLAQ& & & $503$ &$18.00\leq g\leq21.85$& $0.35\leq z\leq 0.75$ & $1.84\pm0.3$& CCF&\cite{2009MNRAS.394.2050M}&\\
        SDSS DR5& & & $2,476$ &$i\leq19.1$& $0.25<z<0.60$ & $1.09\pm0.15$& CCF&\cite{2009MNRAS.397.1862P}&\\
         SDSS DR5& & & 30,239 & $M_{i}<-22$& $0.3\leq z \leq 2.2$ &$2.06\pm0.03$& ACF&\cite{2009ApJ...697.1634R}&\\
        SDSS DR5& & & 7,902 &$M_{i}<-22$& $0.1<z<0.8$&$1.32\pm0.17$& ACF&Shen et al. (2009)&\\
        SDSS DR5& & & 9,975 & $M_{i}<-22$ & $0.8<z<1.4$ &$2.31\pm0.22$& ACF&Shen et al. (2009)&\\
        SDSS DR5& & & 11,304 &$M_{i}<-22$ & $1.4<z<2.0$ &$2.96\pm0.26$& ACF&Shen et al. (2009)&\\
        SDSS DR5& & & 3,828 & $M_{i}<-22$ & $2.0<z<2.5$ &$4.69\pm0.70$& ACF&Shen et al. (2009)&\\
        SDSS DR5& & & 2,693 &  $M_{i}<-22$& $2.9<z<3.5$ &$7.76\pm1.44$& ACF&Shen et al. (2009)&\\
        SDSS DR5& & & 1,788 &  $M_{i}<-22$ & $3.5<z<5.0$ &$12.96\pm2.09$& ACF&\cite{2009ApJ...697.1656S}&\\
        SDSS DR5& & & 1,788 & $M_{i}<-22$ & $3.5<z<5.0$ &$9.85\pm2.27$& ACF&\cite{2009ApJ...697.1656S}&\\
        SDSS DR7& & & 37,290 & $i\leq19.1$ & $0.8\leq z \leq2.2$ &$1.50\pm0.37$& ACF&\cite{2010MNRAS.409.1691I}&\\
         Bo$\ddot{o}$tes$^{n}$& & & 445 &  log $L_{\rm bol}$  $= 45.86$ (median) & $0.7<z<1.8$ &$2.17\pm0.55$& CCF&\cite{2011ApJ...731..117H}&\\
        Bo$\ddot{o}$tes& & & 445 & log $L_{\rm bol}$  $= 45.86$ (median)  & $0.7<z<1.8$ &$2.50\pm0.65$& ACF&\cite{2011ApJ...731..117H}&\\
                RASS$^{o}$& & & 1,552 & $43.7<$ log $L_{\rm X}$ (0.1$-$2.4keV) $< $44.7 & $0.16<z<0.36$ &$1.30\pm0.09$& CCF&\cite{2011ApJ...726...83M}&\\
RASS& & & $629$ & $43.05\lesssim$ log $L_{\rm X}$ (0.1$-$2.4keV) $ \lesssim 44.12$& $0.07<z<0.16$ &$1.19^{+0.08}_{-0.09}$& ACF&\cite{2012ApJ...746....1K}&\\
RASS& & & $1,552$ & $43.69\lesssim$ log $L_{\rm X}$ (0.1$-$2.4keV) $\lesssim44.68 $& $0.16<z<0.36$ & $1.06^{+0.09}_{-0.11}$& ACF&\cite{2012ApJ...746....1K}&\\
RASS& & & $876$ & $44.25\lesssim$ log $L_{\rm X}$ (0.1$-$2.4keV) $\lesssim 45.04$& $0.36<z<0.50$ & $0.96^{+0.22}_{-0.54}$& ACF&\cite{2012ApJ...746....1K}&\\
 XMM/SDSS& & & 297 &$41.0<$ log $L_{X}$ (2$-$10keV)   $<44.0$ & $0.03<z<0.2$ &$1.23^{+0.12}_{-0.17}$& CCF&\cite{2012MNRAS.420..514M}&\\
                $\rm Many^{p}$& & & 1,466 &log $L_{\rm X}$ (0.5$-$8keV)  $\geq41.0 $  & $0<z<3.0$ &$2.26\pm0.16$& ACF&\cite{2013MNRAS.428.1382K}&\\
                 SDSS DR7& & & 8,198 &  $-28.693<M_{i'}<-22.576$ & $0.3<z<0.9$ &$1.38\pm0.10$& CCF&\cite{2013ApJ...778...98S}&\\
                \hline \\
                        \end{tabular}
            }
             \\
    $^{a}$The Great Observatories Origins Deep Survey (GOODS; \citealt{2003mglh.conf..324D}).\\
     $^{b}$The bias factors for quasars which are derived by Francke et al. (2008).\\ 
$^{c}$The 2dF QSO Redshift Survey (2QZ; \citealt{2000MNRAS.317.1014B}).\\
     $^{d}$The total numbers of quasars at $0.80<z<2.10$ (\citealt{2006MNRAS.371.1824P}).\\
     $^{e}$The absolute magnitude range of quasars at $0.80<z<2.10$ (\citealt{2006MNRAS.371.1824P}).\\
     $^{f}$The 2dF-SDSS LRG and QSO (2SLAQ; \citealt{2006MNRAS.372..425C,2009MNRAS.392...19C}).\\
     $^{g}$The number of 2QZ quasar sample.\\
$^{h}$The magnitude range of 2QZ quasar sample.\\
$^{i}$The number of 2SLAQ quasar sample.\\
$^{j}$The magnitude range of 2SLAQ quasar sample.\\
$^{k}$The Multiwavelength Survey by Yale-Chile (MUSYC; \citealt{2006ApJS..162....1G}) Extended Chandra Deep Field-South (ECDF-S; \citealt{2005ApJS..161...21L,2006AJ....131.2373V}).\\
     $^{l}$All-Wavelength Extended Groth Strip International Survey (AEGIS; \citealt{2007ApJ...660L...1D}).\\
     $^{m}$XMM-Newton wide-field survey in the COSMOS field (XMM-COSMOS; \citealt{2007ApJS..172...29H,2007ApJS..172..341C}).\\
     $^{n}$Bo$\ddot{o}$tes multiwavelength survey (Bo$\ddot{o}$tes; \citealt{2007ApJ...671.1365H}).\\
     $^{o}$The ROSAT All Sky Survey (RASS; \citealt{1999A&A...349..389V}).\\
      $^{p}$Chandra Deep Field (CDF)-North, CDF-South, AEIGS, COSMOS, and Extend CDF-South.\\
\end{table*}
\begin{figure}[!t]
\begin{center}
\vspace{-2cm}
\includegraphics[bb= 50 20 350 468,clip,width=8.6cm]{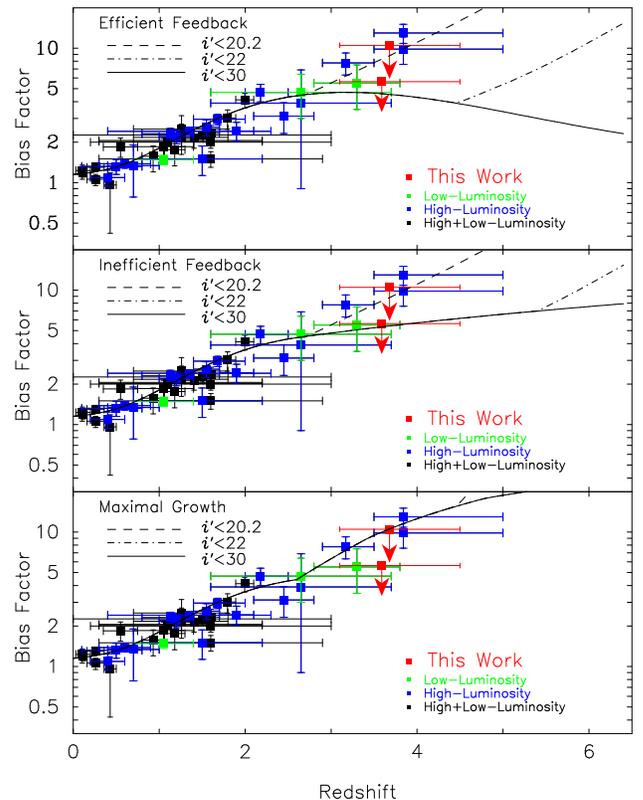}
\caption{Quasar bias factor as a function of redshift. Red, green, blue and black squares show the bias factors for our results (total and spectroscopic sample), low-luminosity quasars ($M_{\rm 1450}>-24$), high-luminosity quasars ($M_{\rm 1450}<-24$), and high+low-luminosity quasars, respectively. Black lines show the quasar bias factors as a function of redshift with three different models. The model shown in the top panel assumes that SMBH growth shuts down after quasar phase (efficient feedback). The model shown in the middle panel assumes that all $z>2$ black holes grow with the observed quasar luminosity function to the characteristic peak luminosities at $z\sim2$, then shut down (inefficient feedback). The model shown in the bottom panel assumes that quasar growth tracks host halo growth, even after a quasar episode, until $z = 2$ (maximal growth). Dashed, Dash-dotted, and solid lines in each panel show the $i'$-band magnitude ranges at $i'<20.2$, $i'<22$, and $i'<30$, respectively (Hopkins et al. 2007).}
\end{center}
\end{figure}

\section{Discussion}
We calculated the quasar-LBG two-point CCF and the LBG ACF at $z\sim4$ to investigate the luminosity dependence of quasar clustering and constrain the dark matter halo mass in which low-luminosity quasars at $z\sim4$ reside.
Since the LBG ACF at $z\sim4$ has been studied by many investigators (\citealt{2001ApJ...558L..83O,2004ApJ...611..685O,2005ApJ...635L.117O,2005MNRAS.360.1244A,2006ApJ...637..631K,2006ApJ...642...63L,2009A&A...498..725H,2011ApJ...737...92S,2014arXiv1407.7316B}), we compare the previous results of $b_{\rm LBG}$ with the derived $b_{\rm LBG}$ in this work (Figure 8). 
 Our inferred bias factor for LBGs ($b_{\rm LBG}$ is $4.92^{+2.07}_{-1.29}$ at $M_{\rm UV}<-20.9$) is not inconsistent with the luminosity dependence of the galaxies bias factor which are reported from previous studies.
We also found that the quasar-LBG CCF shows similar clustering with the LBG ACF while the errors of our results are large, due to the low numbers of low-luminosity quasars.

To constrain the triggering mechanism of the activity in low-luminosity quasars at $z\sim4$, we calculated the bias factor for low-luminosity quasars  in the COSMOS field, which are fainter than the characteristic absolute magnitude\footnote{The characteristic absolute magnitude is the absolute magnitude where the QLF changes its slope from steep at the brighter side to shallow at the fainter side, that is seen typically at $M_{\rm 1450}\sim-24$ (see e.g., Ikeda et al. 2011).}. The $86\%$ upper limits of bias factors of $z\sim4$ low-luminosity quasars are $5.63$  and $10.50$ for the total and the spectroscopic samples, respectively. 
The inferred bias factor for the total sample is smaller than 
   that for the spectroscopic sample, while both results are 
   just the $86\%$ upper limits. In order to clarify the accurate
   bias factor for low-luminosity quasars, we need larger 
   samples of low-luminosity quasars.

The bias factors for low-luminosity quasars derived in this work and results for different redshift and/or luminosity given in previous studies (Adelberger \& Steidel 2005; Croom et al. 2005; Myers et al. 2006; Porciani \& Norberg 2006; \citealt{2008MNRAS.383..565D}; Francke et al. 2008; Gilli et al. 2009; Shen et al. 2009; Ross et al. 2009; Coil et al. 2009; \citealt{2009MNRAS.394.2050M}; \citealt{2009MNRAS.397.1862P}; Ivachenko et al. 2010; Hickox et al. 2011; Miyaji et al. 2011; Krumpe et al. 2012; \citealt{2012MNRAS.420..514M}; Koutoulidis et al. 2013) are showed in Figure 9. We also summarized our result and the previous results of the bias factor for quasars in Table 7. The bias factors which are plotted in Figure 9 and Table 7 are showing results calculated by the ACF or CCF. In order to study the redshift dependence of the bias factor for low-luminosity quasars from $z\sim3$ to $z\sim4$, we compare previous results at $z\sim3$ with our result at $z\sim4$. Francke et al. (2008) calculated the bias factor for AGNs at $z\sim3$ at the UV magnitude range between $-26$ and $-20$ and the obtained bias factor is $5.5\pm2$. Our result for the total sample is consistent with that of their study, though we can not rule out the possibility of a lower value of the bias factor for low-luminosity quasars. We also compare the bias factor for luminous quasars which are brighter than the characteristic absolute magnitude at a similar redshift (Shen et al. 2009). Shen et al. (2009) calculated the bias factor for luminous quasars at $3.5<z<5.0$ and the obtained bias factors are 12.96$\pm2.09$ (excluding negative data points of the correlation function) and 9.85$\pm2.27$ (including negative data points of the correlation function), respectively. Our result for spectroscopic sample is consistent with their study, though we can not rule out the possibility of a lower value of the bias factor for low-luminosity quasars. Our result for total sample is much smaller than that of Shen et al. (2009). The $86\%$ upper limits of the inferred bias factors for low-luminosity quasars at $z\sim4$ correspond to the typical dark matter halo mass are log $(M_{\rm DM}/(h^{-1} M_{\odot}))= 12.7$ and $13.5$ for the total and the spectroscopic samples, respectively. This result is not inconsistent with the predicted bias for quasars which is estimated by the major merger models (e.g., \citealt{2007ApJ...662..110H}).

\cite{2007ApJ...662..110H} predicted the bias factors as a function of redshift using three different models. The first one is that SMBH growth shuts down after each quasar phase (efficient feedback). This model is assuming that each SMBH only experiences one phase of quasar activity and SMBH growth will stop after this one quasar phase. The second one is that all SMBHs at $z>2$ grow with the QLF to the characteristic peak luminosities at $z\sim2$ and after that SMBH growth shut down (inefficient feedback). This model is assuming that $z\sim6$ quasars grow either continuously or episodically with their host systems until $z\sim2$. Hence a quasar feedback at $z>2$ is insufficient to shut down a quasar. The last one is that SMBH growth with a Eddington rate until $z = 2$ (maximal growth). This model is assuming that SMBHs will grow at their Eddington rate until $z\sim2$. 
 In case of efficient feedback and inefficient feedback model,
they predict that the bias factor for low-luminosity quasars becomes
weak with decreasing UV luminosity. In contrast,
Incase of the maximal growth model, this predict that the 
luminosity dependence of the quasar bias is not detected. Our result for total sample is lower than that of the maximal growth model.
However those three models cannot be discriminated by our result for spectroscopic sample. While we can also constrain the quasar lifetime using the quasar bias factor and quasar space density in principle, it is currently too challenging to
constrain it due to the lack of the spectroscopic sample of low-luminosity quasars.
 In order to measure the bias factor for low-luminosity quasars with smaller error bars, we need to use larger samples of low-luminosity quasars. Further observations of low-luminosity quasars in a wider survey area are crucial to provide firm constraints 
 on different scenarios of quasar evolution
 and elucidate the triggering mechanism of low-luminosity quasars, especially at $z>3$, with smaller statistical errors.
 Surveys for high-$z$ low-luminosity quasars with the next-generation wide-field prime-focus camera for the Subaru Telescope (Hyper Suprime-Cam: \citealt{2006SPIE.6269E...9M,2012SPIE.8446E..0ZM}), Euclid (\citealt{2011arXiv1110.3193L}), and the Large Synoptic Survey Telescope (\citealt{2008arXiv0805.2366I}) will address these issues in the very near future.

\section{Summary}
 We have estimated  the quasar-LBG two-point CCF for low-luminosity ($-24<M_{\rm 1450}<-22$) quasars and LBGs at 
$3.1<z<4.5$  in the COSMOS field. Our quasar sample consists of 25 quasars with spectroscopic or photometric redshifts. This sample 
is referred as the total sample. We also use the 16 quasars with spectroscopic redshifts (the spectroscopic sample). We use a sample of
835 LBGs with $z'<25.0$ in  the same redshift range.
We have also estimated  the LBG ACF at $z\sim4$ for comparison with the quasar-LBG CCF at $z\sim4$. Our main results are summarized below.

\begin{enumerate}

\item The correlation amplitudes of the quasar-LBG CCF  are $1.82^{+1.71}_{-0.88}$ and  3.43$^{+3.16}_{-1.64}$ 
for the total and the spectroscopic sample, respectively. The correlation amplitude of the LBG ACF is  3.88$^{+2.37}_{-1.47}$.

\item The $86\%$ upper limits of the spatial correlation lengths for the quasar-LBG CCF  are $7.60$ $h^{-1}$ Mpc and $10.72$ $h^{-1}$ Mpc  for the total and the spectroscopic sample, respectively.
The spatial correlation length for the LBG ACF is 6.52$^{+3.16}_{-1.96}$ $h^{-1}$ Mpc.

\item The $86\%$ upper limits of the bias factors of $z\sim4$ low-luminosity quasars are $5.63$ and $10.50$ for the total and the spectroscopic sample, respectively. The bias factor for the LBGs is 4.92$^{+2.07}_{-1.29}$.

\item We find that the bias factor for spectroscopic confirmed low-luminosity quasars at $z\sim4$ is consistent with the bias factor for luminous quasars at $z\sim4$,
         though we can not rule out the possibility of a lower value of the bias factor for low-luminosity quasars.
 \item We also find that the bias factor for low-luminosity quasars at $z\sim4$ is consistent with that of previous results at $z\sim3$ at similar quasar luminosity,
          though we can not rule out the possibility of a lower value of the bias factor for low-luminosity quasars.
\item The $86\%$ upper limits of the inferred dark matter halo masses are log $(M_{\rm DM}/(h^{-1}M_{\odot}))=12.7$ and $13.5$ for the total and the spectroscopic sample,  respectively. This result is not inconsistent with the predicted bias for quasars which is estimated 
by the major merger models.\\

\end{enumerate}

More specific constraints on SMBH
growth scenarios will be obtained through larger samples of
low-luminosity quasars at high redshifts that will be discovered by
forthcoming wider and deeper quasar surveys.
\acknowledgments

We would like to thank the Subaru staff for their invaluable help and all members of the COSMOS team. We thank Philip Hopkins for providing the data of the predicted bias factor which are calculated by his models. We also thank the referee for useful suggestions and comments that helped us to improve this paper. This work was financially supported in part by the Japan Society for the Promotion of Science (JSPS; Grant Nos. 23244031 and 25707010), and also by the Yamada Science Foundation. TM is supported by UNAM-DGAPA Grant PAPIIT IN104113 and CONACyT Grant Cient\'ifica B\'asica 179662. HI and KM are financially supported by the JSPS through the JSPS Research Fellowship.

\bibliography{adssample}

\end{document}